\documentclass[aps,pre,twocolumn]{revtex4-1}
\usepackage{amssymb}
\usepackage{amsmath,bm}
\usepackage{graphicx,color}
 \usepackage{epstopdf}
\usepackage{epsfig}
\usepackage{undertilde}
\usepackage{multirow}
\usepackage[cal=pxtx]{mathalfa}
\usepackage{accents}
\DeclareMathAccent{\wtilde}{\mathord}{largesymbols}{"65}

\usepackage{stackengine}
\stackMath
\newcommand\tenq[2][1]{%
	\def\useanchorwidth{T}%
	\ifnum#1>1%
	\stackunder[0pt]{\tenq[\numexpr#1-1\relax]{#2}}{\scriptscriptstyle\sim}%
	\else%
	\stackunder[1pt]{#2}{\scriptscriptstyle\sim}%
	\fi%
}

\newcommand{\B}[1]{{\underline{#1}}} 

\begin{document}

\title{Energy dissipation in high speed impact on granular media}
\author{Manish Kumar Mandal$^1$, Saikat Roy$^{1,}$}
\email{Corresponding author: saikat.roy@iitrpr.ac.in}
\affiliation{$^1$ Department of Chemical Engineering, Indian Institute of Technology Ropar, Rupnagar, Punjab, India 140001 }

\begin{abstract}
In this work, we thoroughly investigate the impact process on the granular media in the limit when the ratio of the impact velocity to the acoustic speed becomes of the order of $0.01$-$1$, which is far greater than the existing literature ($0.0001-0.001$). We show that the energy dissipation is largely due to the energy cost associated with the exploration between different metastable states via large scale reorganization of the force chain network. In this regime, the conventional drag force models break down, and the drag force can not be decomposed into a depth dependent static pressure and a depth independent inertial drag as proposed in the existing literature. The high dynamical stress generates acoustic pulses, which propagate longer distances rather than decaying exponentially, as observed in the previous works. In the latter stage of the impact process, the boundary also plays an essential role in the reorganization of the force chains as the reflected acoustic pulses interact with the original impact pulses. Furthermore, we study the scaling of the early stage peak forces with the impact velocity and find that spatial dimensionality strongly influences the scaling. 

\end{abstract}

\maketitle
\section{Introduction}
Granular material is a special class of complex systems composed of many interacting constituents behaving collectively. On top of the inherent complex behaviour of granular material, the response of granular media under high speed impact is notoriously difficult to model analytically since the impact process never reaches a steady state (steady-state velocity different from zero) except when a heavy intruder hits a superlight granular media \cite{pacheco2011infinite}. The existing constitutive laws for the granular materials are applicable only for the steady, fully developed flow conditions, whereas the granular impact process leads to unsteady and complex flow.
 Scarcity of governing equations and insufficient force and trajectory data at the level of grain led to the development of numerous phenomenological models\cite{ciamarra2004dynamics,de2004penetration,lohse2004impact,hou2005dynamics,ambroso2005dynamics,goldman2008scaling} to describe the scaling of crater morphology, collision time, penetration depth with that of impact velocity and grain and intruder properties. 

Following these works, various researchers \cite{allen1957dynamics,katsuragi2007unified,umbanhowar2010granular} proposed that the granular drag force term, $F_d$ can be decomposed into a static depth dependent friction term and a velocity dependent inertial drag term.  Although the drag model was shown to be valid in the past works, but almost all of the experimental and simulational studies in the literature focused on the impact velocities, $V_0$ ($1$ to $5$ $m/s$) that are far below than the velocity scale set by the acoustic speed, $V_a$ ($2000$ to $5000$ $m/s$) in the same media. Recently, \citet{clark2012particle,clark2015nonlinear} cleverly reduced the stiffness of the grains to bring down the force propagation speed and consequently made the low impact velocity approach the force propagation speed. The nature of force propagation was shown to depend on a dimensionless parameter, $B$, which is the ratio of the collision time scale ($t_{col}$) and the time scale set by the intruder impact velocity ($D_p/V_0$), where $D_p$ is the grain diameter and the collision time, $t_{col}$ can be calculated based on the interaction law \cite{clark2015nonlinear,SM}. Impact pulses propagate through the sparse force chains when $B\approx0.1$, whereas $B\to1$ leads to a dense space filling network with a homogeneous front.
Although this study made some interesting observations, but it does not address the real scenario where the high speed impact requires greater amount of energy to be dissipated, and the energy dissipation mechanism can be quite different compared to the low speed impact. Very recent investigation \cite{krizou2020power} suggests some  universal scaling of the early stage peak forces with the impact velocity, and the scaling turns out to be insensitive to the spatial dimension and many other system parameters.  This is a puzzling observation and begs for a detailed study on the nature of the initial forces during an impact in granular media. The process of the granular impact and crater formation has rich physics with wider application in many disciplines like ballistics\cite{forrestal1992penetration,glossner2017instrumented}, astrophysics\cite{robbins2012new}, wind-blown transport of sands via granular splashing \cite{kamath2022scaling} and earth sciences\cite{melosh1989impact}. 
At present, there is very scant literature available on the high speed (comparable to the force propagation speed) impact due to the technological limitations. The nature of the drag forces and the energy dissipation mechanism are completely unknown in this regime. In this work, we employ extensive numerical simulations (both in $2D$ and $3D$) to comprehend the physics of the impact process in the high speed limit. The applicability of the existing drag force models is also tested for a wide range of $V_0/V_a$, from $0.008$ to $0.25$. The mechanism of energy transfer and its eventual dissipation in granular media during high speed impact is unveiled via spatio-temporal monitoring of the displacement field, velocity field and complex force chain networks. The scaling of the early stage peak forces with the impact velocity is thoroughly investigated. Also, the effect of the boundary and its importance in transmitting or holding the
impact stress is explored in detail. 

\section{Simulation methodology}
\subsection{Contact interaction}
Frictional granular material is used as a model system for studying the high-speed impact cratering. Discrete element method (DEM ) simulation is employed to keep track of the particles with frictional interaction taking into account both the normal($F_n$) and the path dependent tangential forces($F_t$).  Simulation is performed both in two and three dimensions. Open source codes, Large-scale Atomic/Molecular Massively Parallel Simulator[LAMMPS] \cite{plimpton1995fast,silbert2001granular,brilliantov1996model} are used and customized to carry out the numerical simulation. In \textit{2D} as well as \textit{3D} the particle-particle interactions are modeled as linear spring-dashpot model with a velocity dependent damping and static friction. We have also used non-linear Hertzian interaction between particles in $3D$. Implementation of static friction is done through the tracking of the elastic part of the shear displacement from the time contact was first made. Particles $i$ and $j$, with position vectors given by ${\B r_i, \B r_j}$, have linear velocities ${\B v_i, \B v_j}$ and angular velocities ${\B \omega_i, \B \omega_j}$ respectively. Grains will experience a normal force, $ \B F^{(n)}_{ij} $ whenever there is a relative normal compression on contact given by $\Delta_{ij}=|\B r_{ij}-D_{ij}|$, where $\B r_{ij}$ denotes the vector joining the centers of mass and $D_{ij}=R_i+R_j$ with $R_i$ and $R_j$ being radii of particles. The normal force is modeled as a Hookean spring like interaction, whereas the tangential force is given by similar linear elastic relation upto the sliding point\cite{bandi2018training} . The force magnitudes are given as,
\begin{equation}
\B F^{(n)}_{ij} = k_n\Delta_{ij} \B n_{ij}-\frac{\gamma_n}{2} \B {v}_{n_{ij}}
\end{equation}
\begin{equation}
\B F^{(t)}_{ij} = -k_t \B t_{ij}-\frac{\gamma_t}{2} \B {v}_{t_{ij}} 
\end{equation}
where $\Delta _{ij}$ and $t_{ij}$ denote normal and tangential displacements respectively; $\B n_{ij}$ denotes the normal unit vector given by $\B r_{ij}/|\B r_{ij}|$.  $k_n$ and $k_t$ are respectively stiffness of the springs for the normal and tangential mode of elastic displacement. For the Hertzian case, contact normal force is given as, $F_{Hertzian}=F_{Hookean}\sqrt{\Delta_{ij}}\sqrt{\frac{R_i R_j}{R_i+R_j}}$. Viscoelastic damping constant for normal and tangential deformation are denoted by $\gamma_n$ and $\gamma_t$ respectively and $\B {v_n}_{ij}$ as well as $\B {v_t}_{ij}$ designate the normal and tangential component of the relative velocity between two grains. The relative normal and tangential velocity are given as:
   \begin{eqnarray}
\B {v}_{n_{ij}}&=& (\B {v}_{ij} .\B n_{ij})\B n_{ij}  \\
\B {v}_{t_{ij}}&=& \B {v}_{ij}-\B {v}_{n_{ij}} - \frac{1}{2}(\B \omega_i + \B \omega_j)\times \B r_{ij}.
\end{eqnarray}
   where $\B {v}_{ij} = \B {v}_{i} - \B {v}_{j}$. Elastic tangential displacement $ \B t_{ij}$ is set to zero when the contact develops for the first time between two particles and is computed using $\frac{d \B t_{ij}}{d t}= \B {v}_{t_{ij}}$ . The simulation also account for the rigid body rotation around the contact point to make sure that $ \B t_{ij}$ always remains in the local tangential plane of the contact. The gravitational forces are also accounted for in the simulation.
The translational and rotational degrees of freedom of the particles are computed using Newton's second law; total forces and torques on particle $i$ are given as:
\begin{eqnarray}
\B F^{(tot)}_{i}&=& m_i\B g +\sum_{j}\left(\B F^{(n)}_{ij} + \B F^{(t)}_{ij}\right) \\
\B \tau ^{(tot)}_{i}&=& -\frac{1}{2}\sum_{j}\B r^{ij} \times \B F^{(t)}_{ij}.
\end{eqnarray}
Note that, the tangential force follows a linear relationship with the relative tangential displacement at the contact point as long as the tangential
force is below the limit set by the Coulomb friction,
   \begin{equation}
   F^{(t)}_{ij} \le \mu F^{(n)}_{ij} \ , \label{Coulomb}
   \end{equation}
where $\mu$ stands for friction coefficient. Upon exceeding this, the contact slips in a dissipative fashion and the tangential displacement is truncated accordingly to satisfy the Coulomb criterion. Simulation also incorporates the effect of inelastic collision for both normal and tangential mode of relative movement via the viscoelastic damping coefficients ($\gamma_{n,t}$) which are related to the coefficient of restitution($\epsilon_{n,t}$) and the collision time as below:

 \begin{equation}
\epsilon_{n,t} = exp(-\gamma_{n,t}t_{col}/2),
\end{equation}
where the collision time $t_{col}$ is given as:

 \begin{equation}
t_{col}  = \pi(2k_n/m- \gamma_{n}^2/4)^{-0.5}
\end{equation}
In our simulation, $\epsilon_{n,t}$ is taken as\cite{silbert2001granular} $0.9$ since the coefficient of restitution for dry sand falls in the similar range. In order to capture the dynamics at the time scale of collision, the time step is set as $t_{col}/50$. $t_{col}$ was calculated for the simulation parameters shown in Table \ref{tt}.
\begin{table}[h]
\caption[Simulation parameters] {Simulation parameters used for both 2D and 3D simulations} \label{tt}
\begin{center}
\begin{tabular}{ccccccc}
\hline
 $k_{n}$ &  $\gamma_{n}$ &  $\frac{k_t}{k_n}$ &  $\frac{\gamma_t}{\gamma_n}$&  $\mu$ & $\epsilon$ & $g$  \\
 \hline
 $2\times10^5$ & $600$ &   $2/7$&  $1$&  $0.5$&   $0.90$ & $9.8$\\
 \hline
\end{tabular}
\end{center}
\end{table}
In Table \ref{tt} parameters are shown only for a reference simulation, we also varied the system parameters for different simulations and describe the same in the text whenever the parameter values change with respect to the reference values. Mass per unit area for the smallest particle (in $2D$ bi-disperse particles are used) $m_g$ is $0.133$. Accordingly, the acoustic speed based on the properties of the smallest diameter particle is approximately $1200$($V_a=\sqrt{\frac{k_n}{m_g}}$). Also, parameter, $B$($=\frac{t_{col}V_0}{D_p}$) can be calculated by estimating the collision time based on the initial impact velocity, $V_0$ and the form of the potential ($F=k_n\Delta^{\beta}$) assuming no viscous dissipation. Here, $D_p$ is the diameter of the smallest particle and $\beta=1$ and $1.5$ for Hookean and Hertzian, respectively. Consequently, $B$ is given as $P(\beta)\left(\frac{V_0}{V_a}\right)^{\left(\frac{2}{\beta+1}\right)}$, where $P(\beta)$ is as follows\cite{clark2015nonlinear},
\begin{equation}
P(\beta)=(\pi(\beta+1)/16)^{\frac{1}{\beta+1}}\frac{4\sqrt{\pi}\Gamma(1+\frac{1}{\beta+1})}{\Gamma(\frac{1}{2}+\frac{1}{\beta+1})}
\end{equation}
For the Hookean case, $P(\beta)=3.9374$. Note that all the parameter values are reported in SI units.

\subsection{Bed preparation}
\begin{figure}[htbp]
  \begin{center}

\includegraphics[scale=0.25]{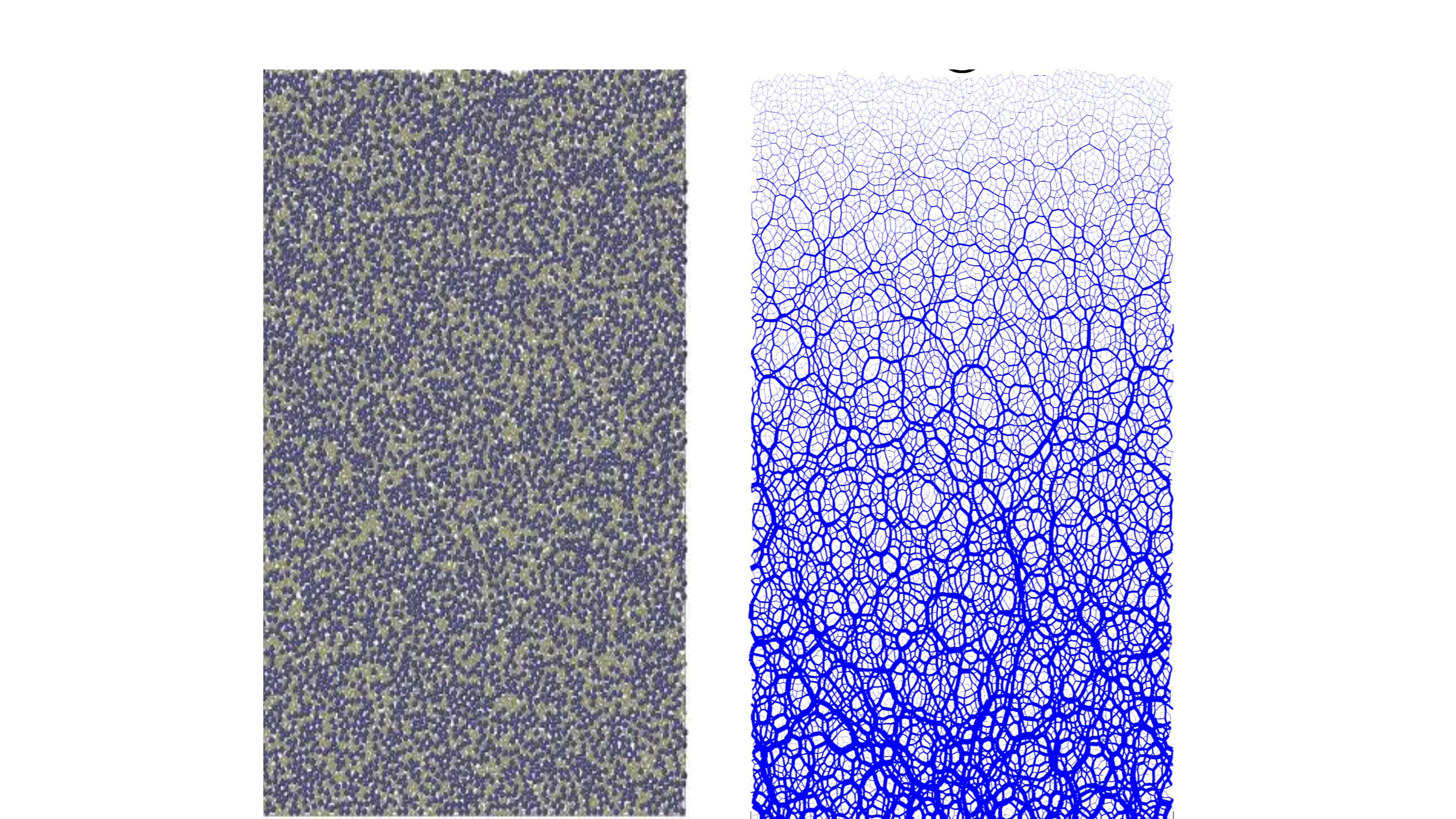}
    \caption{Left: Stable granular bed in two dimension, Right: Corresponding force chain network before the impact}
      \label{fig11int}    
  \end{center}
\end{figure}
For $3D$ simulation of high speed impact, we first make a three dimensional box with periodic boundary conditions in the $x$ and $y$ direction, and a fixed wall at the bottom plane whose outward unit normal is $\widehat{e_z}$. The box has a dimension of $30D_p\times30D_p\times500D_p$.  The length in the $z$ direction is kept long enough to prevent atom loss due to the high speed impact. Next, $N=50000$ (larger system size $N=250000$ is also investigated) mono-disperse spherical particles having diameter, $D_p=0.2 m$ are dropped under gravity and following that, the bed is allowed sufficient time to attain the mechanically stable state with an average force and torque on each particle of the order of $10^{-9}$ with negligible kinetic energy. This stabilization process is extremely important to ensure that we start with a stable bed rather than some fragile configuration which may lead to spurious results. After a stable bed is created, the impacting ball of spherical shape of diameter $10D_p$ is placed very close to the free surface and is given an initial impact velocity in the $z$ direction which is also the direction of the gravity. The average stable bed height is $46D_p$ and the corresponding volume fraction, $\phi$ of the prepared bed is $0.62$, which is very close to the random close packing\cite{scott1969density} state.\\
We also performed simulation in \textit{two dimension} because visualization of the grain scale phenomena is much easier in $2D$ compared to higher dimensions. Similar to $3D$, we first define a simulation box of dimension $100D_{p1}\times 1000D_{p1}$ in the $xy$ plane, followed by pouring of $10000$ (we also simulate a large system,$N=40000$) bi-disperse granular particles, half of which is having diameter, $D_{p1}=0.2m$ and the other half has diameter, $D_{p2}=0.28m$, into the box under gravity. Selection of bi-disperse particles was done to prevent crystallization which is spontaneous for mono-disperse particles in two dimension. The particles were also poured layer by layer instead of pouring them at one go to achieve a stable configuration with sufficient mechanical equilibrium. Pouring all the particle at one go would have caused large collisional stress leading to very long computational time before the system can be relaxed.
Following the similar analysis as done on $3D$ bed, the bed height in $2D$ is calculated to be $139D_{p1}$ having packing fraction of $0.83$. In Fig. \ref{fig11int} , representative schematic of the initial $2D$ granular bed is shown with and without the force chains. The force chain figure is the visual representation of contact forces between particles giving us an idea about the gradient of the stress created due to the gravity. The contact force data was used to draw the force chain of a certain thickness that scales with the magnitude of the force. As expected, the force chains are more dense and thick at the bottom due to the gravitational stress whereas the force chain network is very sparse close to the surface. Force chains represent the stress transmission paths in the granular media and it plays a crucial role in the propagation of any disturbance through the granular materials.

\section{Results and discussions}
\begin{figure*}[htbp]  
\begin{center}
\includegraphics[scale=0.35]{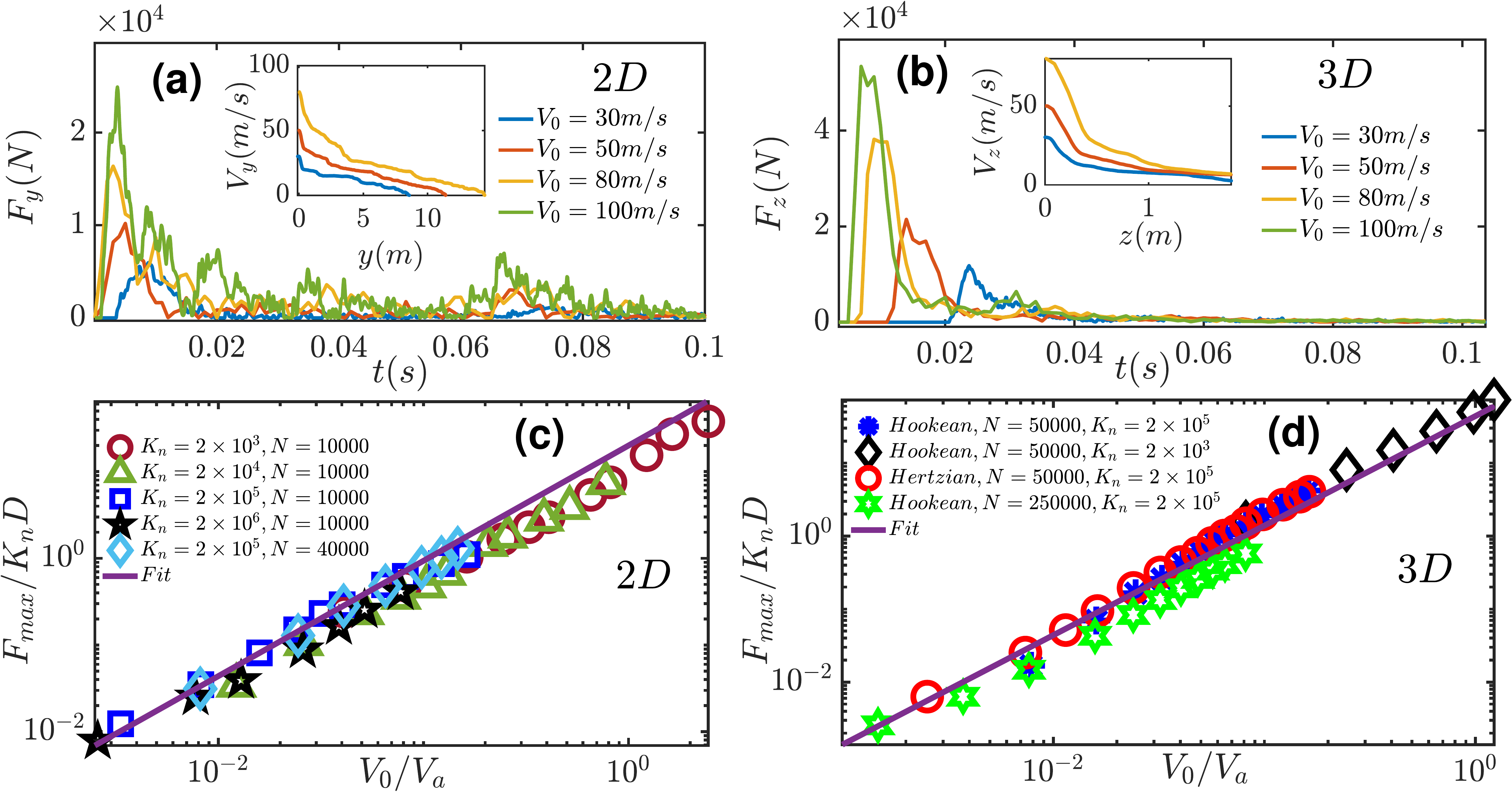}
\includegraphics[scale=0.10]{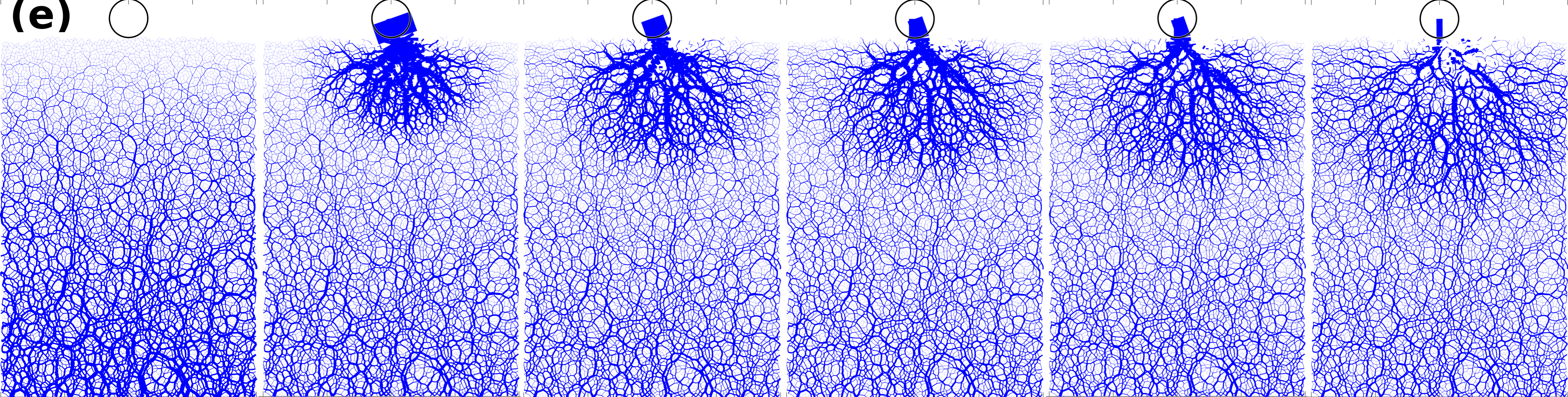}
\includegraphics[scale=0.28]{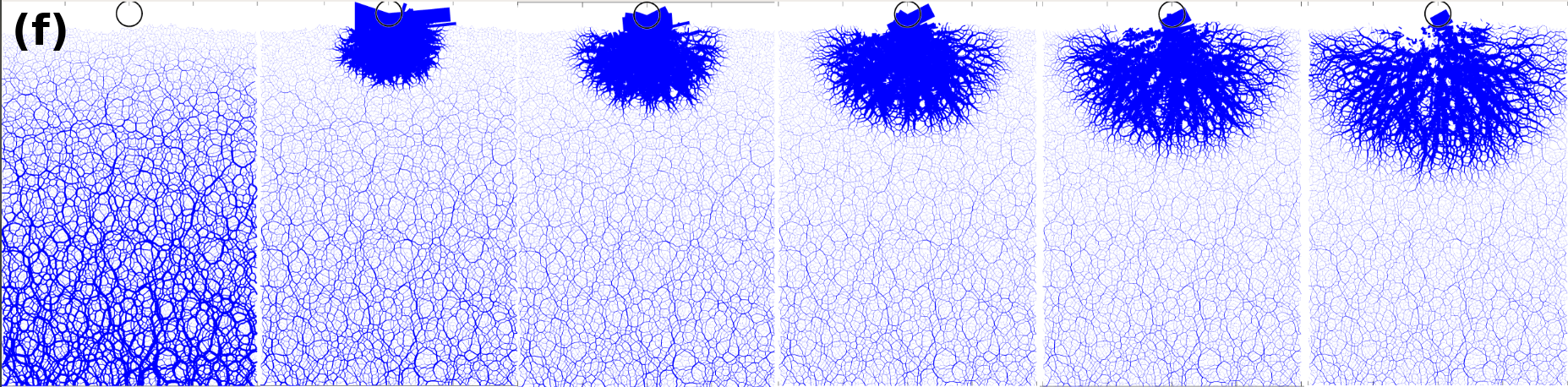}
\caption{(a) Temporal variation of drag force in $2D$ for $V_0/V_a$=$0.03$, $0.04$, $0.06$ and $0.08$: inset shows the corresponding variation of velocity as a function of the intruder depth. (b) Drag force in $3D$ for the similar range of velocities, inset: intruder velocity vs intruder depth. (c) Scaling of the dimensionless peak force, $F_{max}/k_nD$ with the non-dimensional impact velocity, $V_0/V_a$ in $2D$ for different stiffness and system size. The solid line shows a power law fit with an exponent $\sim1.33$. (d) The same scaling in $3D$ for different system size, interaction law and stiffness. The solid fit corresponds to a power law scaling with an exponent $\sim1.5$. (e) and (f) show the force chain evolution (for Hookean interaction) in the initial stages of the impact for $B=0.01$ and $B=0.1$, respectively. Here, the line thickness is scaled according to the contact force magnitude normalized by the mean force. }
\label{fig1} 
\end{center}
\end{figure*}

\begin{figure}[b]
\begin{center}
\includegraphics[scale=0.28]{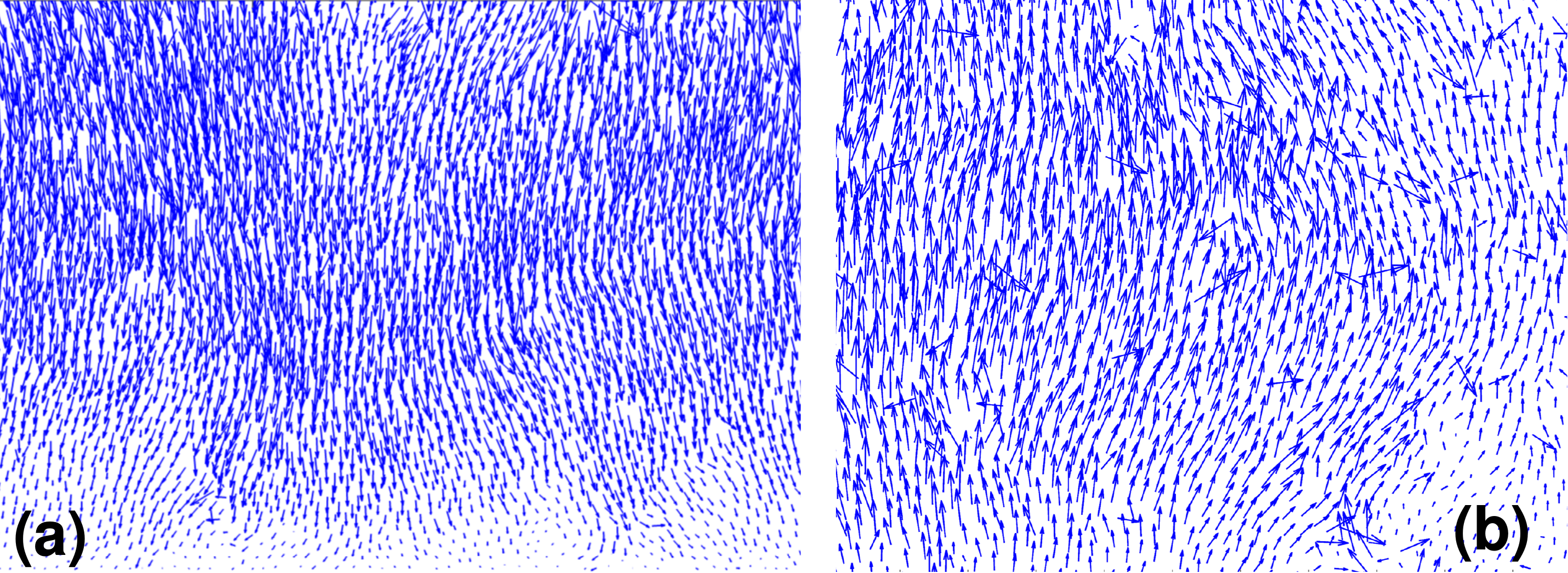}
\includegraphics[scale=0.2]{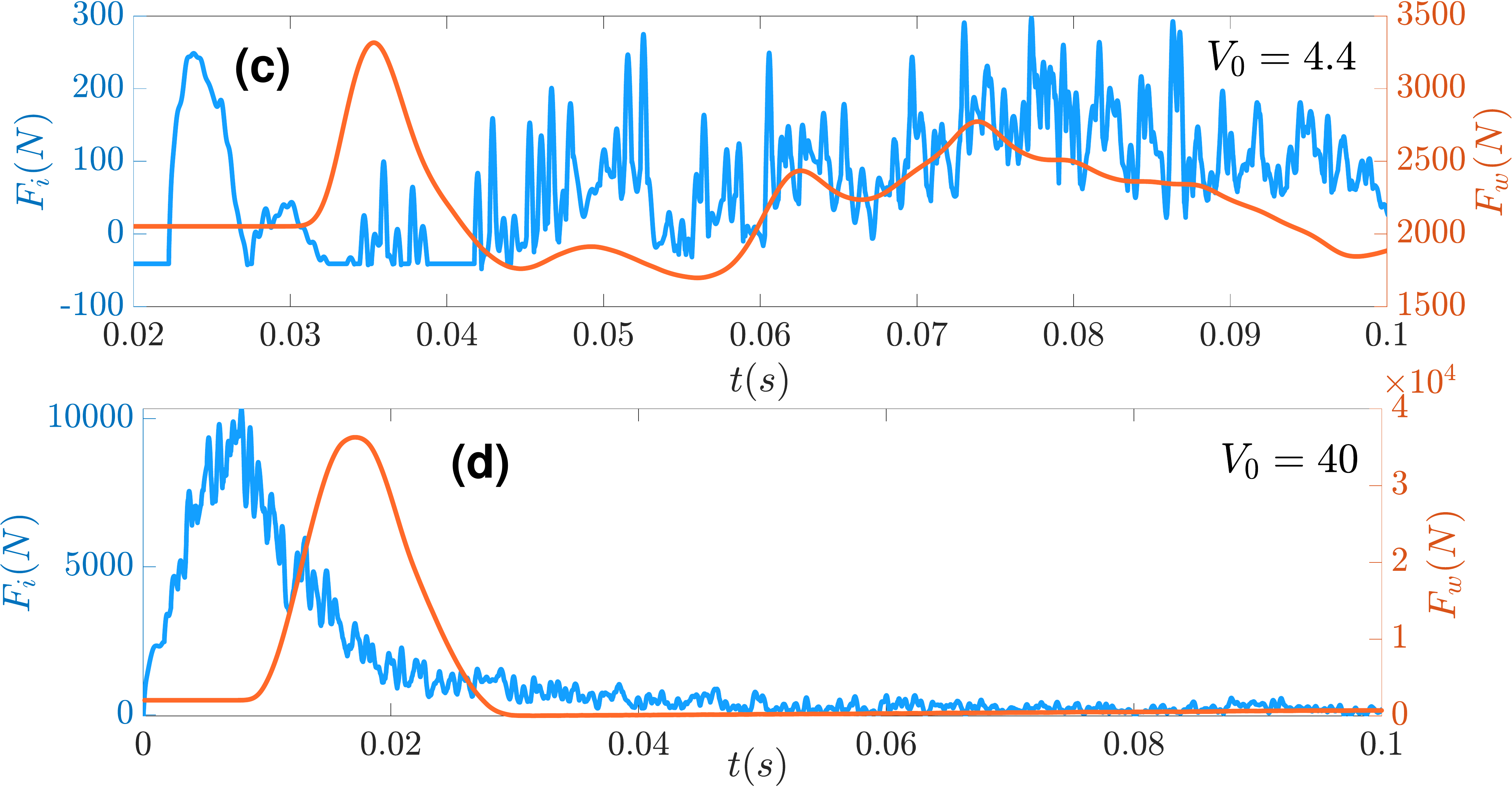}
\caption{(a) Displacement field near the bottom wall showing the arrival of the compression front. Here, $V_0=4.4$ m/s and displacement magnitude is magnified 40 times the original. (b) Elastic unloading of the bed leading to the reversal of the displacement field.(c)-(d) Temporal variation of the force on the wall and the intruder in $3D$ are plotted simultaneously for $V_0=4.4$ and $40$ m/s, respectively. We also observe a similar behavior in $2D$ (not shown).}
      \label{fig2} 
\end{center}
\end{figure}
\subsection{Phenomenology of impact}
We begin our study by analyzing the force on the intruder exerted by the granular media in Fig. \ref{fig1} (a)-(b) for both $2D$ and $3D$. In line with the low speed impact\cite{krizou2020power}, the drag force attains a maximum very quickly, followed by slow relaxations dominated by fluctuations. The fluctuations decay very fast in the three dimension compared to the two dimensional case since the extra dimension gives the system additional direction to relax the effect of the impact.  Also, the response in $3D$ in terms of the peak force becomes stronger than $2D$ with the increase in impact velocity. Fig. \ref{fig1} (c)-(d) depicts the power law scaling of the peak force with $V_0$. Interestingly, the exponent in $3D$ is higher ($\sim1.5$) than in 2D, contradicting the recent results where the scaling exponent ($\sim 1.33$) is independent of the spatial dimension. Since the volume of the configurational space of stress paths (force chains) is greater in $3D$ than $2D$, the number of acoustic events or pulses that carry the intruder energy into the medium also increases proportionately in three dimension compared to the two-dimensional case. Hence, greater resistance to impact is observed in $3D$ than in $2D$ for comparable initial impact speeds. To explicitly show that the observed scaling is not an artifact of the finite size, we vary the system size in both two and three dimension and find that the scaling exponent remains invariant of the system size. We have also checked in all our simulations that the peak force occurs well before the acoustic pulses get reflected from the boundary (Please see supplemental\citep{SM} video 1 showing the pulse propagation in a large system, $N=40000$).
The exponent of the power law scaling is also insensitive to the interaction potential, and the stiffness, in line with the recent observations \cite{krizou2020power}. 

Note that during the attainment of the peak force, the velocity does not reduce significantly and the intruder hardly penetrates. After reaching the peak, the retarding force starts to relax and the ball starts penetrating significantly into the media with ``stop and go'' kind of motion where at some moments the ball is falling freely under gravity even inside the bed and this phenomenon is observed for all ranges of impact speed.  The stronger fluctuations in $2D$ is also reflected in the velocity-depth trajectories (see the inset of Fig. \ref{fig1}(a) and (b)). The shape of the velocity-depth curve is concave upward, which presents striking dissimilarity with the existing literature in the low speed limit \cite{huang2020role,ambroso2005dynamics,umbanhowar2010granular}, where the shape is concave downward and can be reproduced by solving the conventional drag models \cite{huang2020role,katsuragi2007unified,pacheco2011infinite}. This observation is evocative of the possible breakdown of the known macroscopic drag force models in the high speed limit and presents the possibility of unexplored rich grain scale physics.
\subsection{Grain scale picture}
Before we test the drag force models explicitly, we turn our attention to the grain scale picture of the impact process in terms of the spatio-temporal variations of the complex force networks, displacement field and the velocity field. Despite, the recent experiments \cite{clark2012particle,clark2015nonlinear} with photoelastic disks presented some interesting grain scale picture of the impact process, the exact and the complete understanding of the force  network evolution and its effect on the intruder motion is not well understood. Photoelastic measurements have a resolution of $256\times584$ pixels \cite{clark2012particle} at high speed and thus give only a measure of the total photoelastic intensity in an image. In contrast, simulations can provide better insights into the nature of the vectorial contact forces. In Fig.~\ref{fig1} (e)-(f), we show the evolution of the force chain networks in the early stages of the impact for the two values of $B=0.01$ and $B=0.1$. For both cases, before the impact, gravity sets the gradient of the pressure for which the force chains look denser at the bottom. As soon as the intruder strikes the bed, the large dynamical stress dictates the gradient of pressure, and the force chains look denser close to the impact point. The impact energy gets propagated in the form of acoustic pulses, which reach the end of the system boundary even though the system size is large enough to avoid boundary effects (see the Supplemental videos 2-3\cite{SM}). This very fast large length scale propagation of disturbances is evocative of the collective motion of the granular particles, which are correlated upto long range even before the impact.  

We observe the reflection of the acoustic pulses from the boundary and the sideways scattering and branching of the pulses. Reflected pulses also interact with the original pulses emitted from the intruder and give rise to continuous large scale reorganization of the force chains, which results in the temporally fluctuating force on the intruder. None of the existing experimental studies capture such long range propagation of disturbances due to the low resolution of the photoelastic measurements that are typically used to characterize the forces in the experiments. The existing literature, without any physical explanation, suggests an exponential decay of the pulses meaning the pulses decay almost immediately after traveling only a few particle diameter, which is at variance with our simulation observations. We observe that the force propagation happens via a well-defined compression front for $B=0.01$, which is far below that reported ($B=0.6$) in the recent $2D$ experiments \cite{clark2015nonlinear}. Even for $V_0/V_a=0.03$ ($B=0.1$), we see a dense compression front propagating through the media whereas the previous observations showed sparse chain-like force propagation for the same value of $B$. We speculate that the setup used in the experiment had strong side-wall friction and boundary effects which led to the quick damping of the energy pulses. It is also possible that the limited resolution of the photoelastic response at low stress levels makes the determination of the signal propagation far beneath the intruder difficult.

We also simultaneously analyze the particle displacement field near the bottom (see Fig.\ref{fig2} (a)-(b) and also supplemental video 4 \cite{SM}) and observe that a compression front indeed reaches the bottom, and a strong elastic resistance is provided by the bottom wall leading to the flip in the particle displacement field. During the whole process, the particles are moving cooperatively, and the flipping of the displacement field takes longer than the collision time, meaning a large length scale reorganization is inevitable. Intriguingly, the phenomenon of compression and decompression keeps repeating until the intruder comes to rest, which also gives rise to large fluctuations in the force time series. We also show the velocity field at different stages of the impact in the supplemental videos 5-6\cite{SM} for low and high impact velocities, and the long distance propagation of an acoustic pulse is vividly observed. The disturbance propagation speed can also be estimated by monitoring the force on the wall (See Fig.~\ref{fig2} (c)-(d)) and measuring the time taken for the disturbance to reach the wall (time of flight measurement). The force propagation speed is almost of the order of the acoustic speed ($1200 m/s$), and it is independent of the impact velocity for the linear interaction. Surprisingly, the temporal variation of the force on the wall looks very similar to the force-time series of the intruder, albeit with lesser fluctuations since the wall is in contact with large number of force chains. As the wall and the intruder are far apart, a similar temporal response at distant points suggests that the large scale reorganization of the force chain networks dictates the response.  

\subsection{Force network reorganization, dissipation and breakdown of inertial drag models}
\begin{figure}[b]
\begin{center}
\includegraphics[scale=0.17]{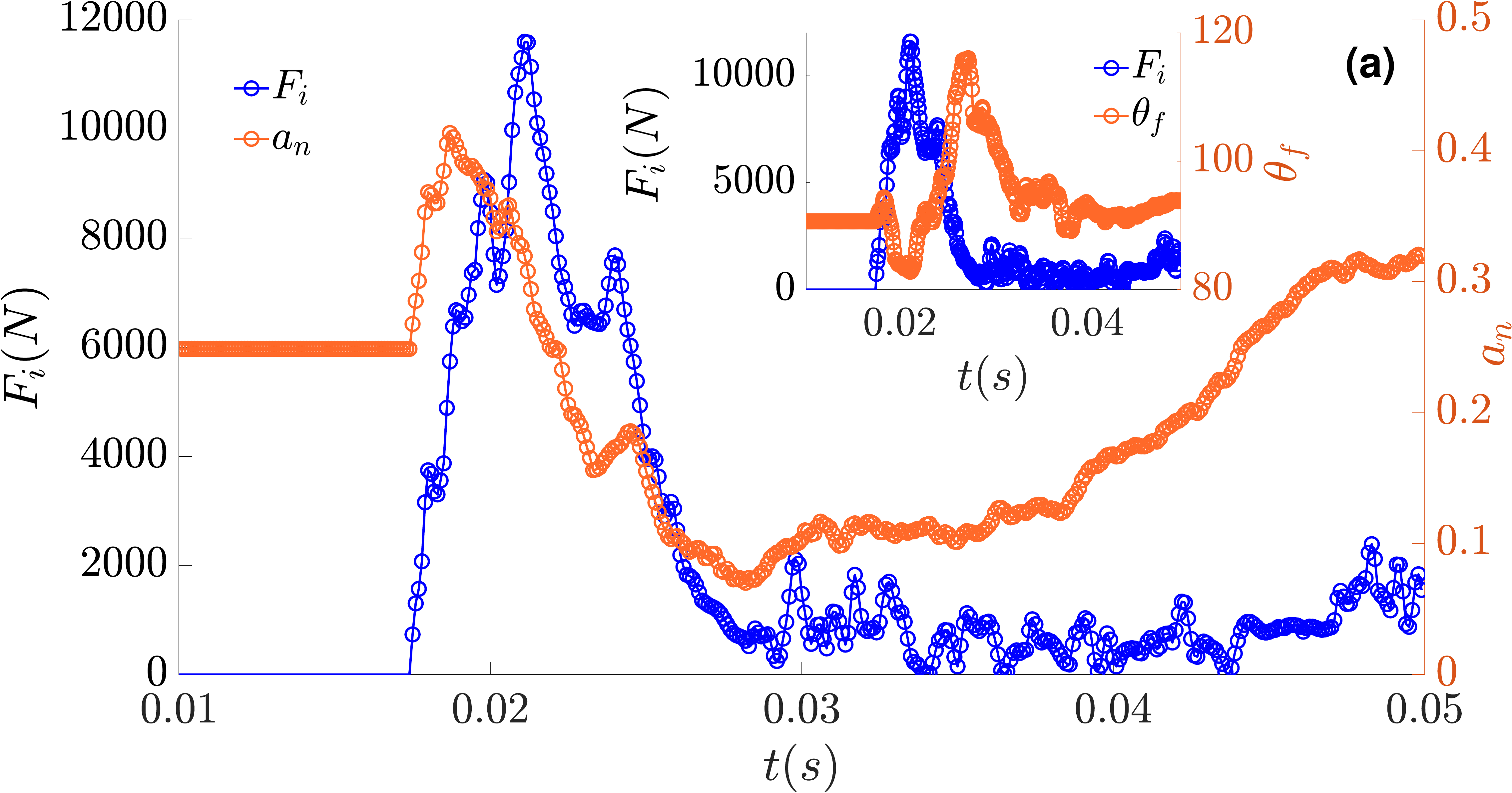}
\includegraphics[scale=0.17]{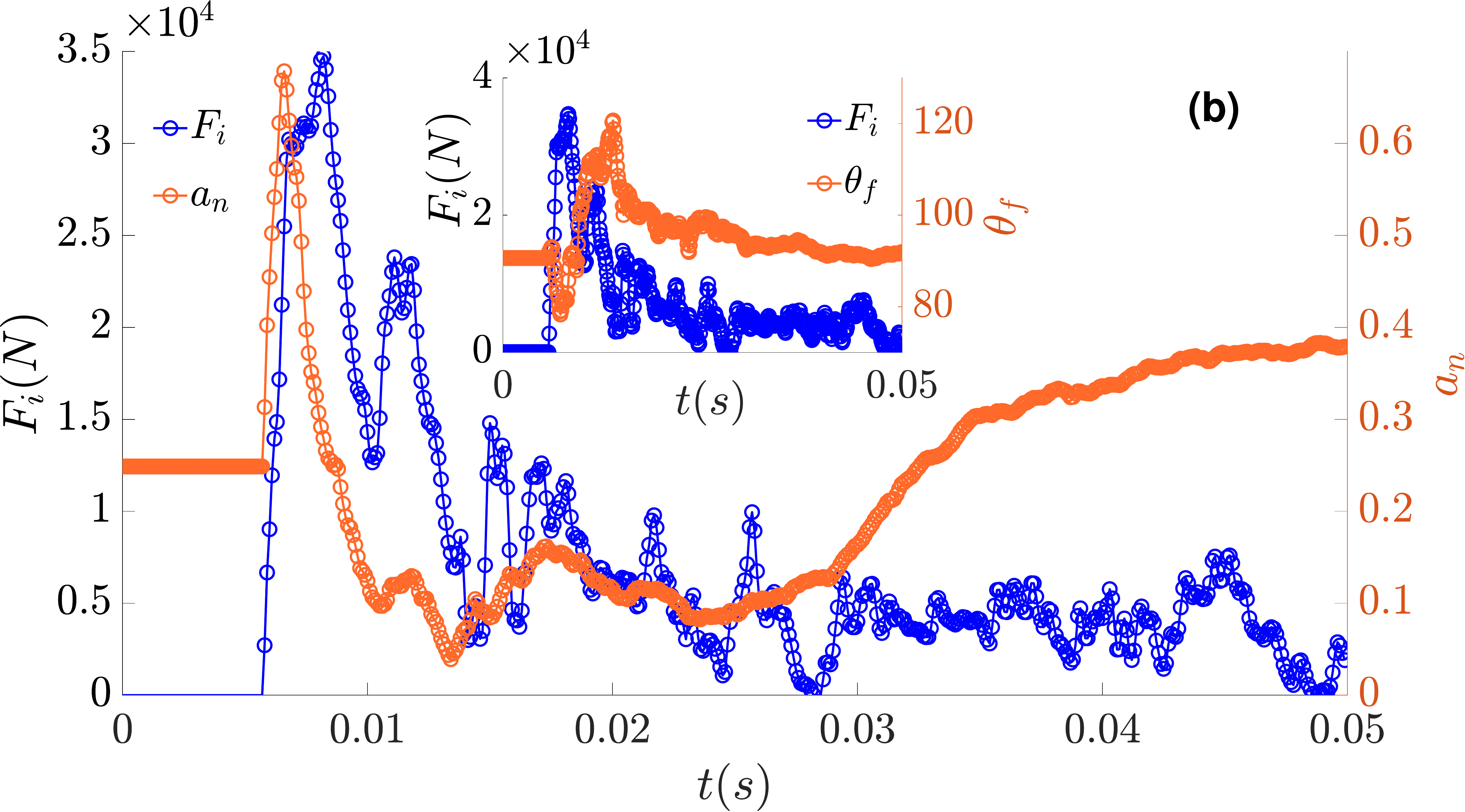}
\caption{(a) Temporal variation of the force anisotropy, $a_n$ (Right $y$-axis) is plotted simultaneously with the temporal variation of the force, $F_i$ on the intruder (Left $y$-axis) for $B=0.2$: The inset shows the change in the orientations of the force network as a function of time (b) Similar to (a) except $B=0.5$. }
\label{fig3} 
\end{center}
\end{figure}
For the quantitative description of the force chain reorganization, we now monitor the anisotropy of the force network and its preferred orientations. As earlier studies showed that friction does not significantly influence the dynamic impact process, we focus on the force skeletons formed by the contact normal forces only. The normal force anisotropy and its preferred direction are defined by $a_n$ and $\theta_f$, respectively. The calculation of these parameters from the discrete simulation data is performed by introducing a second order tensor, $\xi_{ij}\approx\frac{1}{N_g}\sum_{\theta_g} \bar{f_n} n_i n_j$, where $N_g$ denotes the number of orientation intervals spanning from $0$ to $2\pi$, $\theta_g$ is the average orientation of a group and the corresponding average normal force is denoted by $\bar{f_n}$, $n_i$ denotes the Cartesian components of the contact unit normal vector. Anisotropy parameters, $a_n$ and $\theta_f$ are related to the invariants of $\xi_{ij}$ and its principal directions: $a_n=\frac{2\sqrt{(\xi_{11}-\xi_{22})^2+4\xi_{12}^2}}{\xi_{11}+\xi_{22}} ;\; tan\,2\theta_f=\frac{2\xi_{12}}{\xi_{11}-\xi_{22}}$.
\begin{figure}[h]
\begin{center}
\includegraphics[scale=0.18]{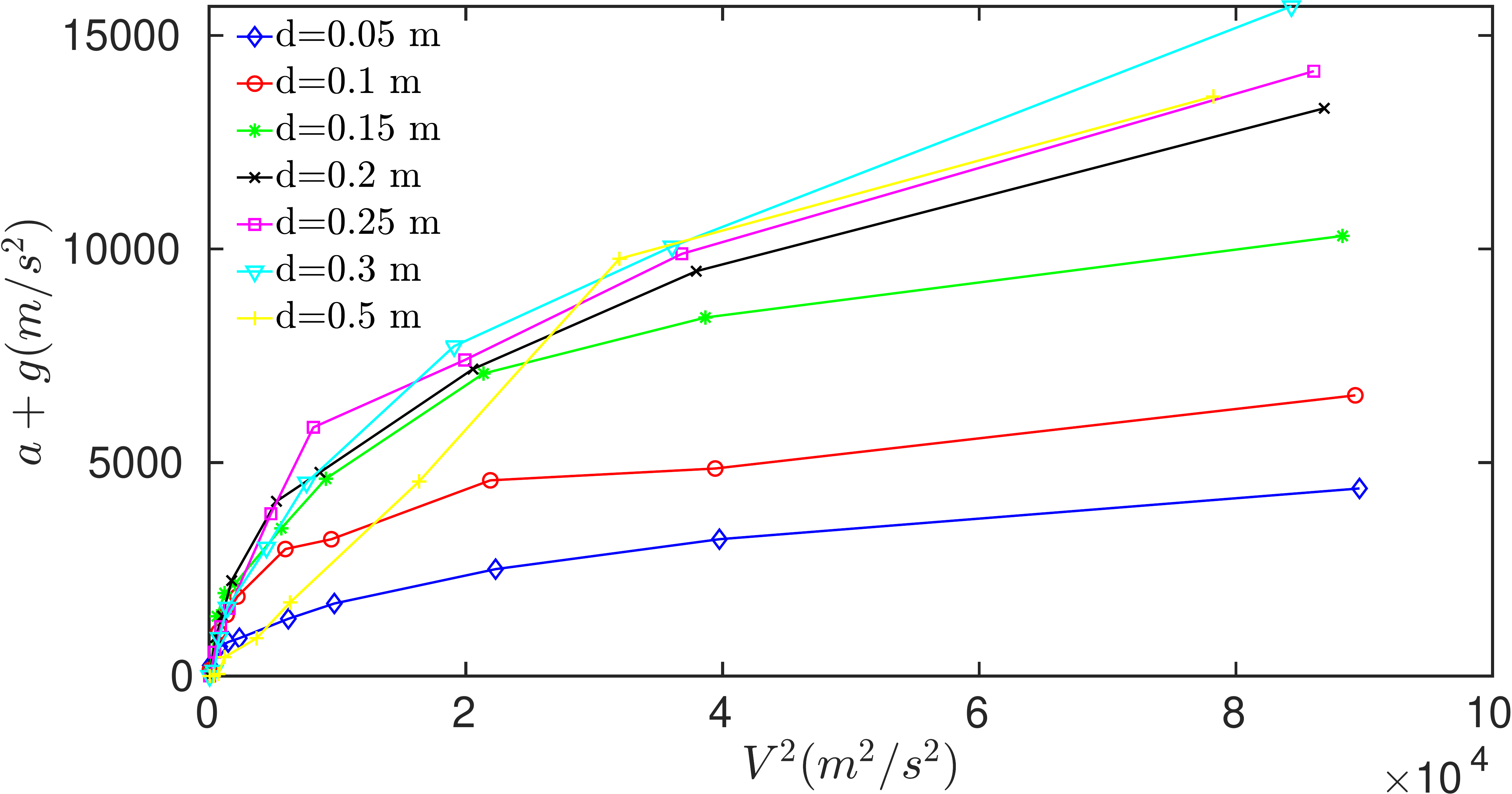}
\includegraphics[scale=0.185]{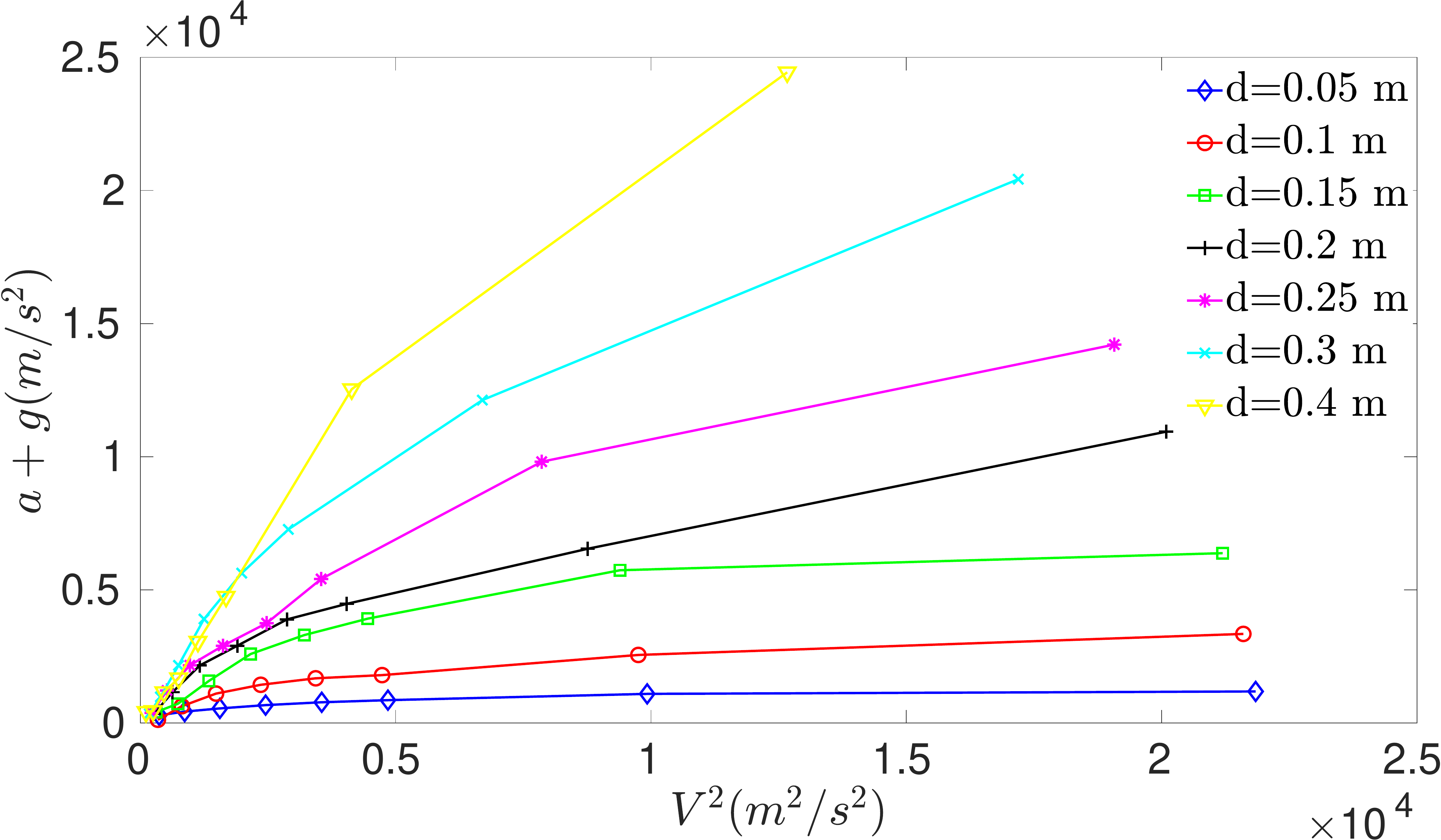}
\caption{(a) Net acceleration, $a+g$ versus square of the velocity, $V^2$ is plotted at seven fixed depths($d$) for different initial impact velocities ($V_0/V_a=0.008$ to $0.25$) in $2D$ , (b) the same is plotted for the three dimensional case for impact velocities ranging from $V_0/V_a=0.02$ to $0.13$.}
\label{fig4} 
\end{center}
\end{figure} 
 Fig.~\ref{fig3} depicts the temporal variation of both the anisotropy and its principal direction, along with the force time series of the intruder for different impact velocities. Before the impact, the force chains are organized mostly in the direction of gravity and $a_n=0.2$. Upon impact, the force chains start to reorganize (see the inset of Fig.~\ref{fig3}) as reflected by the change in the principal direction ($\theta_f$) of the force network. Also, the force network becomes progressively anisotropic to support the sudden impact load and hits a peak, which in turn gives rise to a maximum force on the intruder. After the peak, the force anisotropy decreases in a manner similar to the decrease in the force on the intruder. In the later stages of the relaxation, temporal variation of the force anisotropy decorrelates from the force-time series of the intruder. Furthermore, we find a strong correlation between the orientation of the normal force network and the temporal evolution of the force on the intruder. We also observe a time lag between the force network reorganization and its effect to be felt on the intruder force. The time lag decreases with the impact velocity, suggestive of a decreasing length scale upto which the reorganization occurs. In summary, under high speed impact, the granular media constantly traverses between different fragile states via large scale reorientation of the force networks, and the force on the intruder is the consequence of this large scale reorganization. These transient rearrangements of the force network lead to plastic dissipation and are the principal energy loss mechanism during the high speed impact.
Finally, we test the validity of the existing drag models \cite{katsuragi2007unified,allen1957dynamics} in both $2D$ and $3D$ by monitoring the net acceleration and the speed of the intruder at different fixed depths for different impact velocities. If the depth-independent inertial drag were to apply to our high speed regime, net acceleration would be quadratic in speed resulting in parallel straight lines when $a+g$ is plotted against $V^2$ for different depths. Fig.~\ref{fig4}  instead presents an entirely contrasting picture; depth-dependent quadratic profiles are obtained when net acceleration is plotted against the square of the speed at a fixed depth for different trajectories. Hence, the conventional depth independent inertial drag models are unable to capture the force on the intruder in the high speed limit; rather we observe that the net acceleration varies linearly with the velocity with a depth dependent slope, though this scaling needs to be checked extensively with large data set.
\section{Conclusion}
In summary, we employed large scale numerical simulations to understand the response of the granular media under a high speed impact. Although a large volume of the
work on granular impact exists in the literature, but most of the approaches to tackle such a complex problem are heuristic with insufficient grain scale understanding of the highly dynamic impact phenomenon. This work presents a detailed microscopic length scale picture of the impact process in terms of the evolution of the inhomogeneous force chain networks, displacement field, and the velocity field as the impact progresses. These particle scale information proved to be quite useful in understanding the dissipation mechanism in the granular materials which are neither solid nor fluid. Contrary to the previous works showing the exponential spatial decay of the acoustic pulses, we vividly demonstrate a large-length scale propagation of disturbances that get reflected from the boundary, interfering with the original pulses. These acoustic pulses, in turn, induce large scale reorganization of the force chain network, and the granular media constantly explores different fragile states to support the impact load. Reorientation of the force chains leads to plastic dissipation and the eventual absorption of the impact energy. The large scale temporal evolution of the force chain networks dictates the force on the intruder. Consequently, this novel energy dissipation picture does not corroborate with the conventional drag models and hence, the breakdown of the depth independent inertial drag forces. Furthermore, the power law scaling of the early stage peak forces with the impact velocity shows a dependence on the spatial dimensionality, which is at variance with the past works. The result of this work begs for the development of a novel theoretical framework to explain the drag force on the intruder in the high speed limit. It would also be interesting to study the effect of cohesive interactions on the different aspects of the impact process\cite{ralaiarisoa2022particle} and the scaling of the peak forces since, in a natural setup, attractive forces are expected to be present due to van der Waals forces, humidity, moisture, etc.

\textit{Acknowledgement-}S.R. acknowledges the support of SERB under Grant No. SRG/2020/001943 and the IIT Ropar under ISIRD grant.

\bibliography{ref2}

\end{document}